# Effect of Chemical Doping on the Thermoelectric Properties of FeGa3

N. Haldolaarachchige[1], A.B. Karki[1], W. Adam Phelan[2], Y.M. Xiong[1], R. Jin[1],
Julia Y. Chan[2], S. Stadler[1], and D.P. Young[1*]

[1]*Department of Physics and Astronomy, Louisiana State University, Baton Rouge, Louisiana 70803, USA*
[2]*Department of Chemistry, Louisiana State University, Baton Rouge, Louisiana 70803, USA*
[*]Email: dyoung@phys.lsu.edu

Thermoelectric properties of the chemically-doped intermetallic narrow-band semiconductor FeGa3 are reported. The parent compound shows semiconductor-like behavior with a small band gap ($E_g = 0.2$ eV), a carrier density of $\sim 10^{18}$ cm$^{-3}$ and, a large $n$-type Seebeck coefficient ($S \sim -400$ μV/K) at room temperature. Hall effect measurements indicate that chemical doping significantly increases the carrier density, resulting in a metallic state, while the Seebeck coefficient still remains fairly large ($\sim -150$ μV/K). The largest power factor ($S^2/\rho = 62$ μW/m K$^2$) was observed for Fe$_{0.99}$Co$_{0.01}$(Ga$_{0.997}$Ge$_{0.003}$)$_3$, and its corresponding figure of merit ($ZT = 1.3 \times 10^{-2}$) at 390 K improved by over a factor of 5 from the pure material.

## I. INTRODUCTION

Materials with complex band structures have shown unusual magnetic and transport properties. One example is rare-earth based materials, where the unusual hybridization of $d$- or $f$-orbitals and a broad conduction band forms a narrow gap at the Fermi level.[1-3] Materials[4-8] with the above features tend to have a better thermoelectric performance, since they have a narrow peak in the density of states[9] near the Fermi level. A large Seebeck coefficient is essential in having a high thermoelectric efficiency, which is quantified by the thermoelectric figure of merit: $ZT = \left(S^2/\rho\kappa_T\right)T$, where $S$ is the Seebeck coefficient or thermopower, $T$ is the temperature, $\rho$ is the electrical resistivity, and $\kappa_T$ is the total thermal conductivity ($\kappa_T = \kappa_l + \kappa_e$), where $\kappa_l$ is the lattice or phonon thermal conductivity, and $\kappa_e$ is the electronic thermal conductivity. The best thermoelectric materials[10] have a room-temperature figure of merit of $ZT \sim 1$.

FeGa3 is a narrow-gap diamagnetic[11] intermetallic semiconductor which crystallizes in the tetragonal crystal system adopting the space group $P4_2/mnm$.[12] Density of states (DOS) calculations clearly show the existence of a narrow peak at the Fermi level.[13] Haussermann $et$ $al.$[12]

explained that the DOS is dominated by parabolically distributed, nearly-free electron-like states with $s$-$p$ bands from the Ga network at low energies, but Fe $d$-states hybridize with $p$-states of Ga at higher energies[12], which leads to the narrow band gap ($\sim 0.3$ eV) formation in this compound. High temperature ($313 K < T < 973 K$) thermoelectric properties of FeGa3 were reported by Amagai $et$ $al.$[14], and then more recently, Hadano $et$ $al.$[15] reported a detailed analysis of the thermoelectric properties. However, we have reinvestigated most of the physical properties in detail on FeGa3 from above room temperature down to 10 K, and have investigated the chemical doping effects on its thermoelectric properties.

Chemical substitution, as a means to improving a material's thermoelectric performance, is a logical course to pursue, considering that many doped thermoelectric materials have already shown a promising improvement of their figures of merit.[16, 17] However, some of the chemical doping studies have gauged the improvement of the figure of merit ($ZT$) based solely on an enhancement of the power factor ($S^2/\rho$), which means omitting effects on the thermal conductivity.[18, 19] Here, we report on the synthesis and characterization of pure and chemically-doped FeGa3. Results from measurements of the electrical resistivity ($\rho$),

Seebeck coefficient (*S*), thermal conductivity (*κ*), Hall coefficient, and the calculated thermoelectric figure of merit (*ZT*) are reported.

## II. EXPERIMENTS

Polycrystalline pure and chemically-doped samples were prepared by heating a stoichiometric mixture of starting materials in an alumina crucible inside an rf-induction furnace under a partial pressure of ultra high purity argon gas. After melting, the samples were ground to fine powders, pressed into small pellets, and annealed under vacuum at 800 $^0$C for 48 hours to obtain a homogeneous sample. The crystal structure and phase purity of all the samples were investigated by powder X-ray diffraction using Cu $K_\alpha$ radiation from a Bruker AXS D8 Advance diffractometer. Electrical resistivity (*ρ*) was measured using a standard four-probe method in a Quantum Design Physical Property Measurement System (PPMS) using a bar-shaped sample (1 mm x 1 mm x 2 mm) from 350 K to 3 K. The Seebeck coefficients (*S*) were measured in the PPMS from 350 K to 10 K using a home-built sample holder with a constantan metal standard. Thermal conductivity (*κ*) measurements were performed using a longitudinal steady-state method in the PPMS from 400 K to 200 K. The Hall coefficient (*R_H*) was calculated by measuring the Hall-resistivity at room temperature on a bar-shaped sample in the PPMS in a magnetic field up to 9 Tesla.

## III. RESULTS AND DISCUSSION

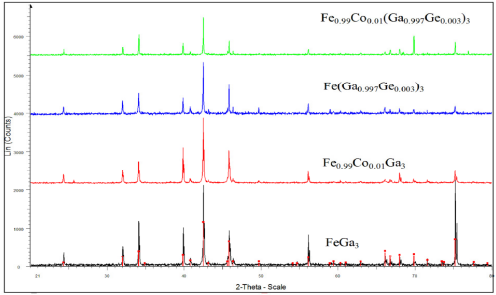

| Compound | a (Å) | c (Å) | Volume (Å³) |
|---|---|---|---|
| FeGa₃ | 6.251 | 6.553 | 256.058 |
| Fe₀.₉₉Co₀.₀₁Ga₃ | 6.258 | 6.553 | 256.632 |
| Fe(Ga₀.₉₉₇Ge₀.₀₀₃)₃ | 6.259 | 6.551 | 256.636 |
| Fe₀.₉₉Co₀.₀₁(Ga₀.₉₉₇Ge₀.₀₀₃)₃ | 6.258 | 6.552 | 256.593 |

Figure 1 (Color online) X-ray diffraction patterns and unit cell parameters of pure and doped polycrystalline FeGa₃. Red dots indicate the standard reference for FeGa₃ compounds.

Standard $2\theta$ X-ray diffraction (XRD) patterns and calculated unit cell parameters of pure and chemically-doped polycrystalline FeGa₃ are presented in Figure 1. The XRD patterns obtained from all the samples indicate homogeneous phases of FeGa₃ without impurity peaks from elemental or secondary phases. Calculated lattice parameters of the pure compound are in good agreement with previously reported data.[12-14] Samples doped with Co or Ge (or both) show changes in unit cell volume compared to that of pure FeGa₃.

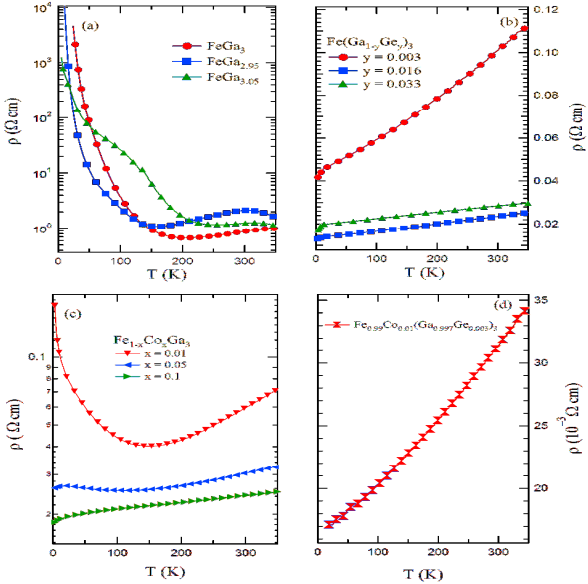

Figure 2 (Color online) Temperature dependent electrical resistivity of (a) pure, off-stoichiomeric, and doped ((b) Co, (c) Ge, (d) Co+Ge) polycrystalline FeGa₃.

Figure 2 shows the temperature dependent electrical resistivity of pure and chemically-doped polycrystalline FeGa₃, as well as the resistivity of off-stoichiometric samples (FeGa₃±ₓ), i.e., samples made with either an excess or deficiency of Ga. Both the pure and off-stoichiometric samples (Figure 2a) show a semiconductor-like behavior, and three regions are distinctively identified in the resistivity of the pure compound (region I: Intrinsic response region above 320 K, region II: Saturation region from 320 K to 140 K, region III: Extrinsic region below 140 K). The band gap (*E_g* = 0.2 eV) was found by fitting the data in region I with an Arrhenius law: $\rho(T) = \rho_0 e^{(E_g/2k_B T)}$, where $\rho$ is the electrical resistivity, $E_g$ is the band gap, $k_B$ is the Boltzman constant, and $T$ is the temperature. The value of the gap obtained from our data is in good agreement with previously reported values[14, 15], but slightly lower than the theoretically estimated value.[13] However, the observed band gap of pure $FeGa_3$ is consistent with a slightly lower resistivity ($\rho(290K)$ = 0.85 $\Omega$ cm) and slightly higher carrier density at room temperature ($n(290K)$ = 2 × $10^{18}$ cm$^{-3}$) than that of the previously reported data for a single crystal.[15] Off-stoichiometric polycrystalline samples ($FeGa_{2.95}$ and $FeGa_{3.05}$) were measured to check the effect of gallium stoichiometry on the resistivity. Both behave like semiconductors, and the room temperature resistivity and calculated band gap of both the gallium deficient and excess samples are similar in magnitude to pure $FeGa_3$.

Chemical substitution, however, has a significant effect on the electrical resistivity (Figure 2b, 2c and 2d). A small percentage of chemical doping on either the iron or the gallium site (1% of Co or 0.3% of Ge) changes the semiconductor-like behavior of the pure compound into metallic, as well as reducing the room temperature electrical resistivity by a large percentage ($\rho(290$ K)$_{Pure\ FeGa3}$ = 0.85 $\Omega$ cm, $\rho(290$ K)$_{Fe0.99Co0.01Ga3}$ = 0.07 $\Omega$ cm, and $\rho(290$ K)$_{Fe(Ga0.997Ge0.003)3}$ = 0.1 $\Omega$ cm). The carrier density at room temperature increases with increasing doping level ($n(290$ K)$_{Fe0.99Co0.01Ga3}$ = 21.5 × $10^{18}$ cm$^{-3}$, and $n(290$ K) $_{Fe(Ga0.997Ge0.003)3}$ = 53.3 × $10^{18}$ cm$^{-3}$), which also provides evidence for changes in the electronic structure of this compound by chemical doping. The electrical resistivity decreases only slightly with further doping over the range of concentrations investigated. Simultaneous doping with both Co and Ge was also investigated for different concentrations. The resistivity of the sample of $Fe_{0.99}Co_{0.01}(Ga_{0.997}Ge_{0.003})_3$ (Figure 2d) has shown a very low electrical resistivity ($\rho(290$ K) = 0.028 $\Omega$ cm) and a higher carrier density ($n(290$ K) = 76.1 × $10^{18}$ cm$^{-3}$) relative to that of the other Co- and Ge-doped samples. This enhances the figure of merit of this sample more than any other of the samples we investigated.

The temperature dependent Seebeck coefficients of the pure and doped samples of $FeGa_3$ are shown in Figure 3. Pure polycrystalline $FeGa_3$ has a large room temperature thermopower ($S$(290 K) = - 440 μV/K) which slowly decreases in magnitude with temperature from 350 K to 100 K and then rapidly decreases with temperature down to 10 K. The negative sign of the thermopower is consistent with the negative Hall coefficient ($R_H$ = - 2.6 × $10^{-6}$ m$^3$/C) at room temperature, which qualitatively agrees with the literature.[15] However, the room temperature thermopower of our sample is slightly larger than the previously reported value for single crystals.[15] Off-stoichiometric $FeGa_3$ shows a slight increment of the room temperature thermopower with gallium excess and a slight decrement for gallium deficiency, but both samples show the same temperature dependent behavior as pure $FeGa_3$, which confirms there is not a considerable effect on the thermopower from slight Ga off-stoichiometry.

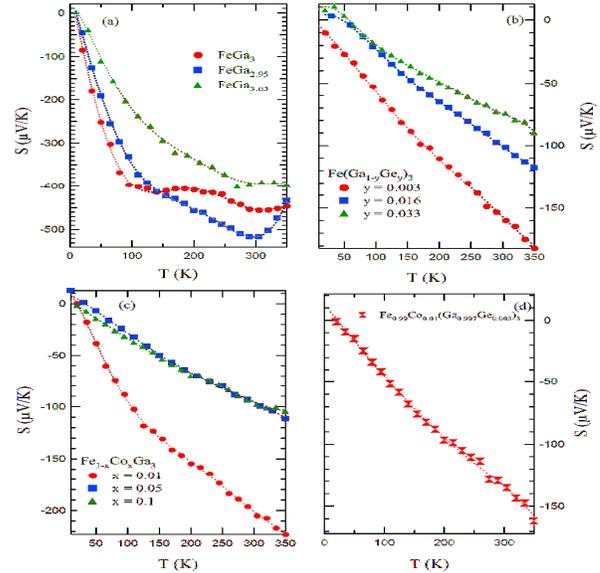

Figure 3 (Color online) Temperature dependent Seebeck coefficient of (a) pure, off-stoichiomeric, and doped ((b) Co, (c) Ge, (d) Co+Ge) polycrystalline $FeGa_3$. (Dotted lines are guide to the eye)

The Seebeck coefficient of $FeGa_3$ is, however, very sensitive to chemical doping (Figure 3b, 3c, and 3d). A small percentage of chemical substitution (1% of Co or 0.3% of Ge) changes the room temperature thermopower significantly. Reduction of the room temperature Seebeck coefficient is consistent with an increase in the carrier density of the system by chemical doping. However, the room temperature thermopower at very low doping concentrations is still considerably large ($S(290\ K)_{Fe0.99Co0.01Ga3}$ = - 190 μV/K, $S(290\ K)_{Fe(Ga0.997Ge0.003)3}$ = - 155 μV/K, and $S(290\ K)_{Fe0.99Co0.01(Ga0.997Ge0.003)3}$ = - 120 μV/K). The negative sign of the thermopower agrees with the negative Hall coefficients ($R_H(290\ K)_{Fe0.99Co0.01Ga3}$ = - 0.29 x $10^{-6}$ m³/C, and $R_H(290\ K)_{Fe(Ga0.997Ge0.003)3}$ = - 0.11 x $10^{-6}$ m³/C), indicating that the majority charge carriers are electrons. Increasing the doping concentration further (1 - 10% of Co or 0.3 - 3% of Ge) slowly reduces the thermopower, and after 10% of Co or 3% of Ge doping, $S(290\ K)$ becomes very small, which results in a low figure of merit.

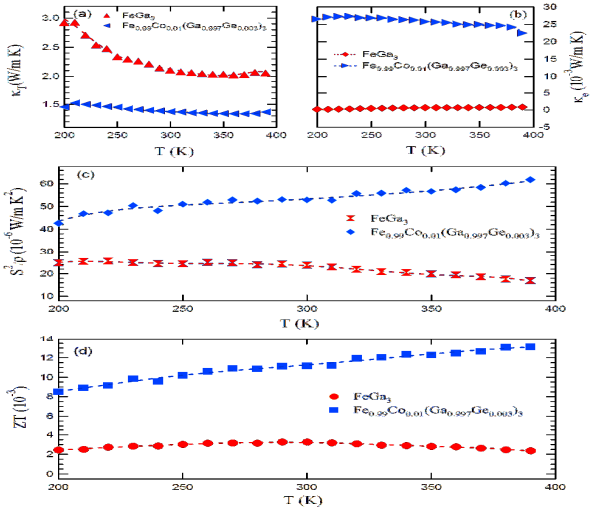

Figure 4 (Color online) Temperature dependent total thermal conductivity (a), electronic thermal conductivity (b), power factor (c) and the thermoelectric figure of merit (d) of pure and doped (Co+Ge) polycrystalline FeGa₃. Electronic thermal conductivity calculated by using the Wiedeman-Franz law. (Dotted lines are guide to the eye)

The temperature dependent thermal conductivity (total thermal conductivity: $\kappa_T$ and electronic thermal conductivity: $\kappa_e$), the calculated power factor, and the calculated thermoelectric figure of merit of pure and doped FeGa₃ are shown in Figure 4. The thermal conductivity of pure FeGa₃ ($\kappa(290\ K)$ = 2.14 W/m K) at room temperature is slightly lower than that of single crystal data.[15] The thermal conductivity gradually increases with decreasing temperature and typically peaks at lower temperature, where the phonon mean free path is comparable to the crystallite size in the sample. The electronic part of the thermal conductivity (Figure 4b) is estimated by the Wiedeman-Franz law ($\kappa_e = L_0T/\rho$), where $L_0$ is the Lorentz number (2.45 × $10^{-8}$ W Ω/$K^2$), $\rho$ is the electrical resistivity, and $T$ is the temperature. The calculated electronic part of the thermal conductivity of FeGa₃ is almost negligible over the entire temperature range when compared with the total thermal conductivity, which agrees well with previously reported data[15], and is consistent with the lattice contribution dominating the thermal conductivity in a semiconducting material. Furthermore, chemical doping at the concentrations presented here does not have a significant effect on the total thermal conductivity of this compound, even though the doped samples show a small change in cell volume and a significant increase in the number of charge carriers. However, simultaneous doping of 1% of Co and 0.3% of Ge shows that the room temperature value of the total thermal conductivity ($\kappa(290\ K)_{Fe0.99Co0.01(Ga0.997Ge0.003)3}$ = 1.4 W/m K) decreased by a factor of 0.6 from that of pure FeGa₃, which can be attributed to enhanced phonon scattering in the system via chemical disorder. The estimated electronic part of the thermal conductivity of the chemically-doped compound has increased by two orders of magnitude over that of pure FeGa₃, which is mainly due to the increased carrier density of the system, but it is still negligible when compared to the lattice part of the thermal conductivity.

The calculated power factor ($S^2/\rho$) shows a large increment (Figure 4c) in the range of temperature from 200 K to 400 K over that of pure FeGa₃. This is primarily because of the significant decrease in the electrical resistivity of the doped samples from that of the pure compound, while maintaining a fairly high thermoelectric power. The effect of the improvement in the power factor can be identified by observing the significant increase in the thermoelectric figure of merit (Figure 4d) of the chemically-doped compound over the whole range of temperature. The largest figure of merit ($ZT$(390 K)$_{Fe0.99Co0.01(Ga0.997Ge0.003)3}$ = $1.3 \times 10^{-2}$) increases by a factor of 5.6 over that of pure FeGa$_3$.

## IV. CONCLUSIONS

In summary, we have synthesized and characterized the thermoelectric properties of pure and chemically-doped FeGa$_3$. The parent compound is a small band-gap semiconductor with a large *n-type* Seebeck coefficient at room temperature, but the power factor and the corresponding figure of merit are small because of a large electrical resistivity. Even small amounts of chemical doping have a significant effect on the physical properties and electronic structure of the pure compound, which shows a large decrease in resistivity with a corresponding increase in carrier density at room temperature. The Seebeck coefficients of the doped samples remain large, with magnitudes greater than 100 µV/K. The highest power factor ($S^2/\rho$(390 K) = 62 µW / K$^2$ m) and figure of merit ($ZT$(390 K) = $1.3 \times 10^{-2}$) at 390K are observed for Fe$_{0.99}$Co$_{0.01}$(Ga$_{0.997}$Ge$_{0.003}$)$_3$. FeGa$_3$ provides an excellent example of a system where chemical doping can be used to optimize the physical properties in the thermoelectric figure of merit and greatly improve the performance.

## V. ACKNOWLEDGEMENTS

DPY acknowledges support from the State of Louisiana Board of Regents (Post Katrina Support Fund, Grant No. 06-719-S4). JYC and WAP acknowledge partial support from NSF-DMR0756281. SS acknowledges support from the National Science Foundation (Grant No. NSF-DMR-0545728).